\documentclass[11pt,twocolumn,superscriptaddress]{revtex4}   
\pdfoutput=1
\usepackage{amsmath}
\usepackage{graphicx}
\usepackage{natbib}
\usepackage{mathrsfs}
\usepackage[T1]{fontenc}
\usepackage[letterpaper,textwidth=7in,top=.75in,bottom=.75in]{geometry}
\def\sbrk#1{{\left[#1\right]}}

\begin{document}
\title{Rubidium dimers in paraffin-coated cells}

\author{V.\ M.\ Acosta}
\email{vmacosta@berkeley.edu}
\address{Department of Physics, University of California,
Berkeley, CA 94720-7300, USA}
\author{A.\ Jarmola}
\address{Laser Centre, University of
Latvia, Rainis Blvd. 19, Riga LV-1586, Latvia}
\author{D.\ Windes}
\address{Department of Physics, University of California,
Berkeley, CA 94720-7300, USA}
\author{E.\ Corsini}
\address{Department of Physics, University of California,
Berkeley, CA 94720-7300, USA}
\author{M.\ P.\ Ledbetter}
\address{Department of Physics, University of California,
Berkeley, CA 94720-7300, USA}
\author{T.\ Karaulanov}
\address{Department of Physics, University of California,
Berkeley, CA 94720-7300, USA}
\author{M.\ Auzinsh}
\address{Laser Centre, University of
Latvia, Rainis Blvd. 19, Riga LV-1586, Latvia}
\author{S.\ A.\ Rangwala}
\address{Light and Matter Physics, Raman Research Institute, C.
V. Raman Avenue, Sadashivanagar, Bangalore 560080, India}
\author{D.\ F.\ Jackson Kimball}
\address{Department of Physics, California State University --
East Bay, Hayward, California 94542-3084, USA}

\author{D.\ Budker}
\email{budker@berkeley.edu}
\address{Department of Physics, University of California,
Berkeley, CA 94720-7300, USA}
\address{Nuclear Science Division,
Lawrence Berkeley National Laboratory, Berkeley CA 94720, USA}

\begin{abstract}Measurements were made to determine the density of rubidium dimer vapor in paraffin-coated cells. The number density of dimers and atoms in similar paraffin-coated and uncoated cells was measured by optical spectroscopy. Due to the relatively low melting point of paraffin, a limited temperature range of $43\mbox{-}80^{\circ}~{\rm C}$ was explored, with the lower end corresponding to a dimer density of less than $10^7~{\rm cm^{-3}}$.  With one-minute integration time, a sensitivity to dimer number density of better than $10^6~{\rm cm^{-3}}$ was achieved.  No significant difference in dimer density was observed between the cells.
\end{abstract}





\maketitle 

\section{INTRODUCTION}
For metrology with atomic spins, the spin-projection-noise-limited sensitivity scales as the square root of the spin relaxation time \cite{Bra1975,Bud2007}. Unfortunately, for thermal vapors in closed containers, collisions with the cell walls typically destroy atomic spin after just one bounce. An early solution to this problem was to fill cells with high densities of a gas with very low polarizability, such as noble gases, which prevent atoms from reaching the cell walls over diffusion-limited times on the order of seconds \cite{Dic1953}. However, in these so-called buffer-gas cells, spin relaxation still occurs via spin-destruction collisions with the noble gas species, and ensemble dephasing can reduce sensitivity when magnetic field gradients are present.

Early work on optical pumping led to the striking realization that if the walls of an alkali-metal vapor cell were coated with paraffin (hydrocarbon chains, ${\rm C}_{n}{\rm H}_{2n+2}$), the atomic polarization relaxed at a much slower rate \cite{Rob1958}. Shortly thereafter, vapor cell technology improved to the point where cells with high-quality paraffin coatings enabled polarized alkali atoms to bounce between the cell walls up to 10,000 times before they depolarized \cite{Bou1966}. These early reports of the remarkable qualities of paraffin have led to widespread application of coated cells in areas where long-lived atomic polarization is desired--for example, in optical magnetometers \cite{Ale2004,Aco2006,Hig2006,Bal2006,Led2007,Bud2007,Was2010}, quantum memories \cite{Bud1998,Jul2004,She2006,Kle2009,Lvo2009}, and atomic clocks \cite{Fru1983,Bud2005}. Measurements with paraffin-coated cells without buffer gas are relatively immune to small magnetic field gradients and currently hold the record for the longest alkali-vapor spin relaxation times--up to one minute for Rb atoms at near-zero magnetic field \cite{Bal2010}.

While paraffin-coated cells have been used for over half a century, the micro-chemistry of the surface coatings has only recently been explored \cite{Sel2010}, and the atomic depolarization mechanisms in these cells \cite{Rob1982,Lib1986,Gra2005} are still not fully understood. In a simple model of an evacuated alkali vapor cell at room temperature, the two dominant sources of depolarization are due to atom-atom ``spin-exchange'' collisions and atom-wall collisions. Recently, evidence for unexplained additional relaxation due to ``electron spin-randomization'' collisions has been observed for Rb, Cs, and K in paraffin-coated vapor cells \cite{Bud2005,Gra2005,Guz2006}. Understanding additional sources of relaxation is especially important in light of the discovery of new cell coatings which operate at higher temperatures (corresponding to higher vapor density) \cite{Sel2008,Sel2009,Ram2009} and the very recent advent of a new coating which enables nearly one million bounces before depolarization \cite{Bal2010}.

In this article we examine whether gaseous impurities, of which homonuclear diatomic molecules (dimers) are the most abundant, are a substantial source of atomic-spin relaxation in paraffin-coated cells. In Ref. \cite{Bud2005}, the ``electron spin-randomization'' collisions were shown to contribute $\sim10~{\rm Hz}$ to the atomic magnetic resonance linewidths at $43^{\circ}~$C. For collisions with dimers to explain such an effect, the dimer density would have to be approximately four orders of magnitude greater than the thermodynamic equilibrium density, assuming an atom-dimer electron spin-randomization cross section on the order of $10^{-14}~{\rm cm^{-2}}$ \cite{Kad2001}. As dimer formation involves a three-body collision most likely to occur near a surface (see, for example, Ref. \cite{ClaMcC64}), hypotheses were formulated that paraffin might catalyze formation of dimers resulting in vapor densities greatly exceeding those expected for a non-interacting vapor in thermodynamic equilibrium \cite{Oku2006}.

We measured the relative dimer and atom densities in both coated and uncoated cells in order to test for these possible effects. The detection of dimers in paraffin-coated cells presents a significant technical challenge, as the melting temperature of paraffin is $\lesssim 80^{\circ}~$C \cite{Rah1987}, corresponding to a dimer density of $\lesssim5\times10^8~{\rm cm^{-3}}$ (typical molecular spectroscopy experiments are performed in the range $250\mbox{-}400^{\circ}~$C where dimer number densities are between $10^{13}$ and $10^{15}$ $\mbox{cm}^{3}$ \cite{Ami1997,Set2000,Ban2005}). At such low temperatures, optical absorption is highly suppressed, as the dimer population is spread over many ro-vibrational levels, so we used fluorescence to measure dimer density. By using frequency-modulation techniques we achieved a sensitivity to dimer number densities of $\sim10^6~{\rm cm^{-3}}$ after one minute of averaging, and were therefore able to operate at low enough temperatures to avoid melting the coating.

\section{Theory}
The law of mass action, when applied to a gas of element, X, in a closed cell reveals that the thermodynamic equilibrium condition is $N_{\rm X}^{2}/N_{\rm X_2}=Z_{\rm X}^{2}/Z_{\rm X_{2}}$ (see, for example, Ref. \cite{GreenBook2}), where $N$ is the total number of gas-phase particles and $Z$ is the single-particle partition function. If we can assume that the particles do not interact with each other or the cell walls, then, for alkali atoms in the ground state (electronic angular momentum $J=1/2$), the sum over all momentum and internal states gives a number-density ratio of:
\begin{equation}
\frac{\sbrk{\rm X}^2}{\sbrk{\rm X_2}} \approx (\frac{\pi m k_B T}{h^2})^{3/2} \frac{(4 I+2)^2}{F(T)},
\label{eq:dimers1}
\end{equation}
where $m$ is the atomic mass, $k_{B}$ is the Boltzmann constant, $T$ is the ambient temperature, $I$ is the nuclear spin, and the brackets refer to number density. $F(T)$ is the sum over internal dimer states given by:
\begin{multline}
F(T)=\frac{e^{\frac{D_0}{k_B T}}}{1-e^{-\frac{\omega_e}{k_B T}}} (2I+1) \\
\times \left( \begin{gathered}
I \displaystyle\sum_{J=0}^\infty(2J+1)e^{-\frac{B_e J(J+1)}{k_B T}} \\
+\displaystyle\sum_{J'=0}^\infty(4J'+3)e^{-\frac{B_e (2J'+1)(2J'+3)}{k_B T}}
\end{gathered} \right),
\label{eq:dimers2}
\end{multline}
where $B_{e}$ is the rotational constant, $\omega_{e}$ is the vibrational constant, and $D_{0}$ is the dissociation energy for the ground electronic molecular state \cite{GreenBook2}. This ratio depends only on spectroscopic constants and temperature, and is therefore a cell-geometry invariant measure which can be compared to the observed vapor density.

\section{EXPERIMENT}
Three cylindrical (1.7 cm length, 2.0 cm diameter) isotopically-enriched (more than $90\%$) $^{87}$Rb vapor cells were used, labeled P1, B1, and P2 (the P stands for paraffin and B for buffer gas). Cells P1 and P2 were paraffin-coated cells prepared following the procedure described in Ref. \cite{Ale2002}, while B1 was an uncoated cell containing 3 torr of Ne buffer gas. To check that the cell coatings were of high quality, spin relaxation measurements were performed using a modified ``Relaxation in the dark'' technique \cite{Fra1959} outlined in Ref. \cite{Gra2005}. These measurements revealed spin relaxation times of $97(13)~{\rm ms}$, $94(11)~{\rm ms}$, and $5(2)~{\rm ms}$ for P1, P2, and B1 respectively \cite{dimernote}, which are within the range of typical values for paraffin-coated and low pressure buffer-gas cells \cite{Fru1985,Bud2007}.

\begin{figure}
\includegraphics[width=.45\textwidth]{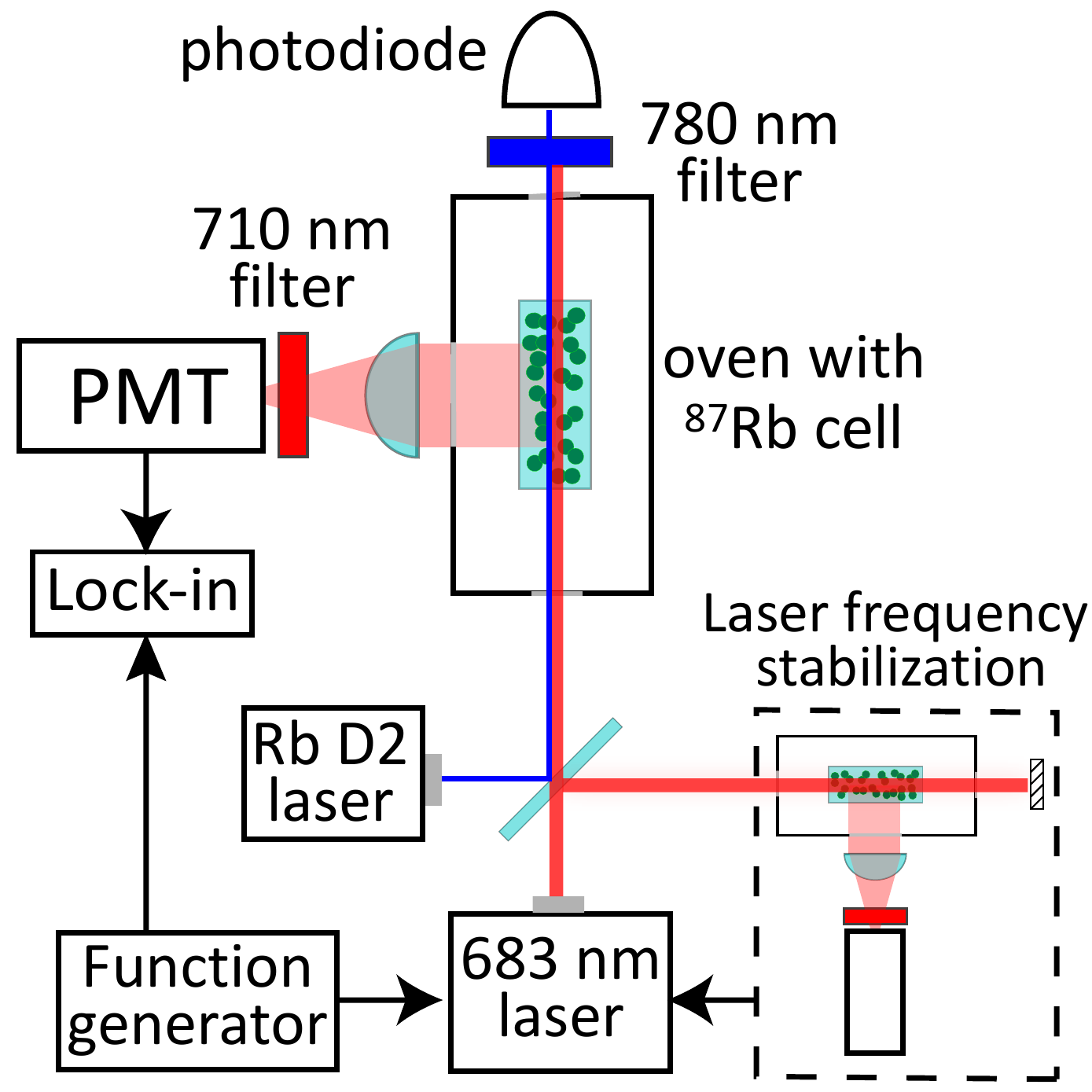}
    \caption{\footnotesize Experimental apparatus used for fluorescence detection of dimers. Using frequency-modulation techniques, the number density of dimers and atoms in cells housed in an oven with optical access was measured by exciting with light from a 683 nm laser and collecting fluorescence around 710 nm. Fluorescence from an uncoated cell in a second, nearly identical oven operating at $150^{\circ}~{\rm C}$ was used to stabilize the laser frequency. The atomic densities were measured by detecting the D2 absorption spectrum using a 780 nm laser. PMT - photomultiplier tube.}
    \label{fig:apparatus}
\end{figure}

The apparatus is shown in Fig. \ref{fig:apparatus}. To monitor dimer density, light from a wavelength-stabilized $683\mbox{-}{\rm nm}$ laser diode with a volume holographic grating was split by a beam splitter and directed into the heated vapor cells, exciting the Rb$_2$ ${\rm X}~^1{\Sigma_{g}}\rightarrow{\rm B}~{^1\Pi_{u}}$ electronic transition. The excitation wavelength was chosen because it is near the band head of the ${\rm X}~^1{\Sigma_{g}}\rightarrow{\rm B}~{^1\Pi_{u}}$ transition, and other optical transitions, for example to the ${\rm A}~{^1\Sigma_{u}}$ excited state, can be neglected. For all measurements, the laser intensity was $\sim0.1~{\rm W/cm^2}$, chosen to maximize fluorescence intensity while remaining in the linear absorption regime. The laser current was modulated to produce sinusoidal frequency oscillations with a modulation depth of $\sim500~{\rm MHz}$ (on the order of the Doppler broadened linewidths) and modulation frequency of 90 kHz. The central frequency of the laser was scanned by a slower current ramp over a range of $\sim 5$ GHz ($\sim0.2~{\rm cm^{-1}}$) at a repetition rate of $100~{\rm Hz}$. The fluorescence from each cell was collected by a lens with a collection efficiency (percentage of total fluorescence) of $\sim0.5\%$, spectrally filtered with an interference filter, and detected with a red-sensitive photomultiplier tube. The interference filter was tilted to transmit light at $710~{\rm nm}$ (full width, half maximum (FWHM) 12 nm), which maximized the transmitted fluorescence intensity. The resulting spectra are referred to as excitation spectra throughout this manuscript. The excitation spectrum from a reference cell maintained at $150^{\circ}~{\rm C}$ was used to stabilize the laser frequency against drifts. Both signals were demodulated with lock-in amplifiers and averaged for approximately one minute with an oscilloscope. As the resulting spectra are approximately the frequency derivatives of the unmodulated spectra, they are referred to as derivative spectra here.

For temperature-dependence studies, an oven, consisting of two high-temperature heat-tape spools and isolating cell mount, heated the cells with a variable equilibrium temperature in the range $43\mbox{-}140^{\circ}~{\rm C}$. Two thermocouples placed at different locations near the surface of the cell monitored the temperature. The atomic vapor density was monitored by analyzing absorption of light from a $780~{\rm nm}$ laser which scanned across the Rb D2 line. The collection efficiency for each cell was periodically measured by simultaneously recording the D2 absorption and fluorescence spectra at low temperature, using the same optical path and collection geometry as for dimer fluorescence.

\section{Results}

\begin{figure}
\includegraphics[width=.45\textwidth]{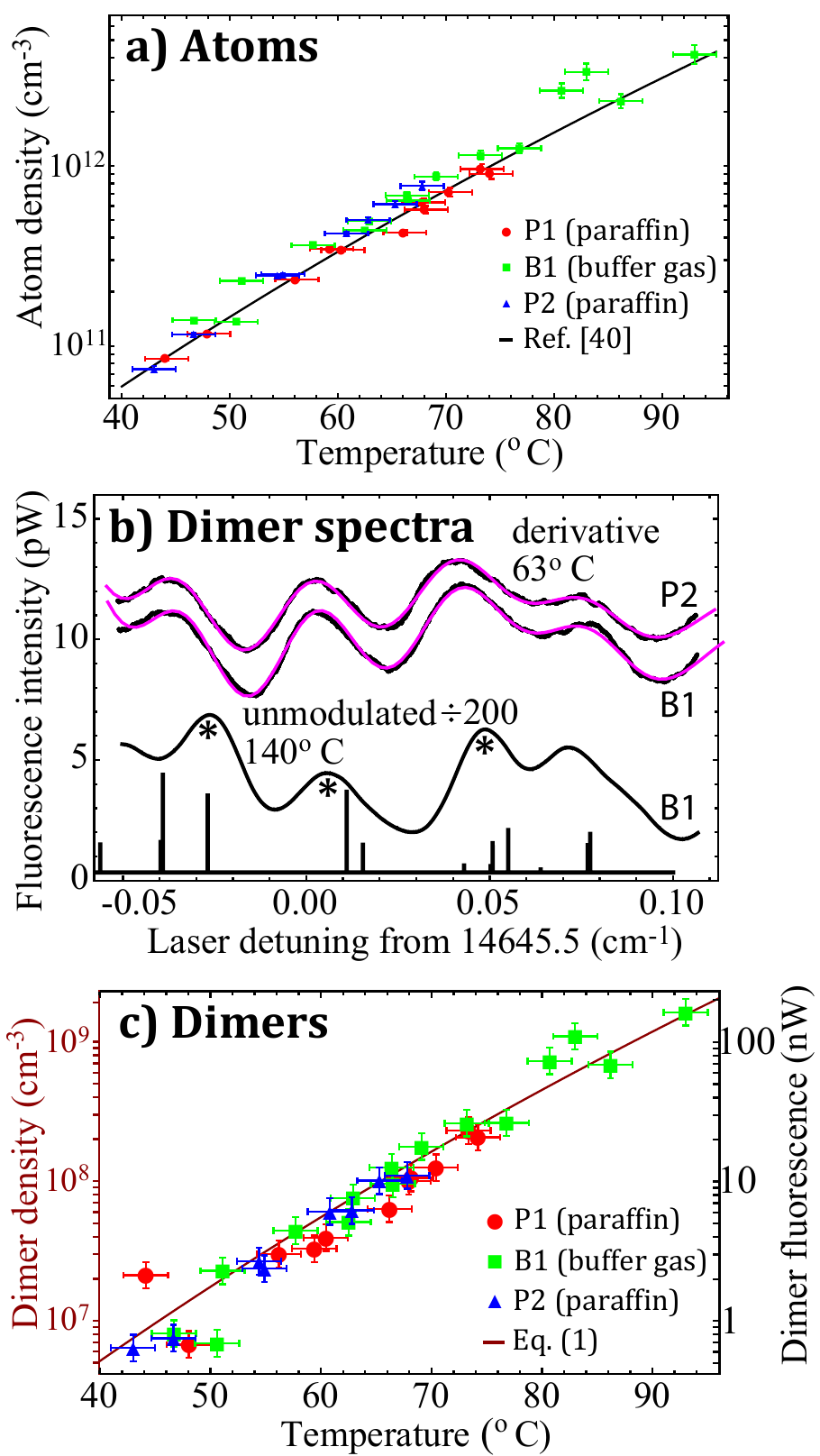}
    \caption{\footnotesize Results of dimer and atom number density measurements. a) Atom number density as a function of temperature (as measured by a thermocouple) for each cell along with empirical number density from Ref. \cite{Crc2007}. Horizontal error bars are estimated based on thermal gradients in the oven and vertical error bars take into account uncertainties in the Doppler width and isotope fraction as well as standard error in the fit. b) Dimer excitation spectra after one minute of averaging in cells B1 and P2. The derivative signals at $63^{\circ}~$C obtained from the lock-in amplifier for both cells (black points, offset from zero for visual clarity) are plotted alongside fits of the phenomenological function described in the text (magenta). The unmodulated spectrum for B1 at $140^{\circ}~$C is also shown, divided by 200 for visual clarity. Asterisks label the peaks used to determine number density. The calculated transition frequencies and amplitudes are illustrated with black lines at the bottom. c) Dimer fluorescence amplitude (normalized as discussed in the text) as a function of temperature for all cells along with the theoretical number density from Eq. \eqref{eq:dimers1} (scaled to fit B1). The horizontal error bars are the same as in (a) and vertical error bars are largely dominated by the uncertainty in fluorescence collection efficiency, with the remainder coming from statistical uncertainty from the fits.}
    \label{fig:alldata}
\end{figure}

Figure \ref{fig:alldata}(a) shows the atomic number density as a function of temperature for the uncoated and two paraffin coated cells. The atomic number density was determined by fitting the absorption spectra with a linear absorption model. The experimental values are consistent with the empirical formulas in Ref. \cite{Crc2007} to within the error bars, which are dominated by uncertainty in temperature measurement.

Figure \ref{fig:alldata}(b) shows an example of both unmodulated and derivative dimer excitation spectra around $14645.5~{\rm cm^{-1}}$ for paraffin-coated cell P2 and uncoated cell B1 along with a calculated excitation spectrum. The transition frequencies and Franck-Condon factors were calculated using spectroscopic data from Refs. \cite{Ami1997,Set2000}. The excitation probability was calculated by weighting the Franck-Condon factors for each allowed ${\rm X}~^1{\Sigma_{g}}\rightarrow{\rm B}~{^1\Pi_{u}}$ vibrational transitions \cite{Ami1997,Set2000} by the thermal equilibrium occupancy of the relevant ro-vibrational states in the ground electronic state, while the emission probability was determined by weighting the vibrational Franck-Condon factors for the relevant excited-state vibrational levels by the spectral profile of the interference filter. As seen in Fig. \ref{fig:alldata}(b), the unmodulated excitation spectrum is consistent with the calculated spectrum to within the $\sim.01~{\rm cm^{-1}}$ precision of the calculated transition frequencies \cite{Ami1997}.

For extracting the dimer density, each derivative spectrum was fit with an empirical function of five Gaussian derivatives. Figure \ref{fig:alldata}(c) depicts the dimer number density versus temperature for each cell. The mean of the amplitudes of each of the three central peaks (labeled with asterisks in Fig. \ref{fig:alldata}(b)) was normalized by the collection efficiency for the respective cell to give the overall fluorescence amplitude at each temperature. This fluorescence amplitude is directly proportional to the dimer density. The absolute dimer number density was estimated by assuming the vapor in the uncoated cell, B1, obeys the thermodynamic equilibrium condition given by the law of mass action, and fitting Eq. \eqref{eq:dimers1} to the B1 values to obtain the scaling factor. The data from all three cells agree with the model for thermodynamic equilibrium to within the uncertainty ($\sim10^6~{\rm cm^{-3}}$ for the lowest temperatures studied here). 

\section{Conclusion}
 The atom and dimer number densities in the uncoated and two paraffin-coated cells over the temperature range examined in this work are consistent with the thermodynamic equilibrium values. At $\sim43^{\circ}~$C, the observed differences in dimer number densities are smaller than $\sim10^6~{\rm cm^{-3}}$, which means that dimer densities in paraffin-coated cells are not a significant source of atomic spin relaxation. Even at the paraffin melting temperature of $\sim80^{\circ}~$C, the broadening of magnetic resonances due to dimer-atom collisions is at the negligibly low level of tens of mHz. The technique developed in this work is well suited for studying light desorption of dimers from the cell coating, an effect which has been known to dramatically increase atomic vapor densities \cite{Goz1993,Mar1994,Ale2002,Gra2005,Kar2009}.

\section{Acknowledgments}
This work is supported by ONR MURI grant \# N-00014-05-1-0406, NSF grants PHY-0855552 and PHY-0652824, and by an NSF/DST Indo-US Collaboration grant. The authors are grateful to A. I. Okunevich, M. Tamanis, O. Nikolayeva, K. Ravi, A. Sharma, and J. Higbie for useful discussions, and to B. P. Das for his support of the project.

\bibliographystyle{apsrev}

\begin{thebibliography}{43}
\bibitem[{dim()}]{dimernote}
\bibinfo{note}{Using this technique, there are actually two time constants
  which characterize spin relaxation (see, for example, Ref. [20]). For
  simplicity, we have quoted the longer of the two times which is, for the
  cells studied in this work, approximately two times longer than the shorter
  time constant.}
\end{thebibliography}

\end{document}